\documentclass{aa}
\begin{document}
\title{A Photometric Catalogue of Southern Emission-Line Stars
\thanks{Based on observations collected at the European Southern Observatory, 
La Silla, Chile and on observations collected at the South African Astronomical 
Observatory.  
Tables 2--4 are only available in electronic form at the CDS via 
anonymous ftp to cdsarc.u-strasbg.fr (130.79.128.5) or via 
http://cdsweb.u-strasbg.fr/cgi-bin/qcat?J/A+A/(vol)/(page).}}

\author{D. de Winter\inst{1,2,3} \and M.E. van den Ancker\inst{4} \and 
        A. Maira\inst{1} \and 
        P.S. Th\'e\inst{5} \and H.R.E. Tjin A Djie\inst{5} \and 
        I. Redondo\inst{1} \and C. Eiroa\inst{1} \and
        F.J. Molster\inst{5,6}}
\institute{
Dpto. F\'{\i}sica Te\'orica, C-XI, 
 Universidad Aut\'onoma de Madrid, Cantoblanco, 28049 Madrid, Spain \and
Instituto de Astrof\'{\i}sica de Canarias, 
 C/ Via L\'{a}ctea s/n, 38200 La Laguna, Tenerife, Spain \and
TNO TPD, Stieltjesweg 1, P.O. Box 155, 2600 AD  Delft, The Netherlands \and
Harvard-Smithsonian Center for Astrophysics, 60 Garden Street, MS 78, 
 Cambridge  MA 02138, USA \and
Astronomical Institute, University of Amsterdam, 
 Kruislaan 403, 1098 SJ  Amsterdam, The Netherlands \and
School of Materials Science and Engineering, Georgia Tech, 
 Atlanta  GA 30332-0245, USA
}

\offprints{Dolf de Winter (dolf@xiada.ft.uam.es)}
\date{Received $<$date$>$; accepted $<$date$>$} 
 
\abstract{We present a catalogue of previously unpublished optical 
and infrared photometry for a sample of 162 emission-line objects 
and shell stars visible from the southern hemisphere.  The data 
were obtained between 1978 and 1997 in the Walraven ($WULBV$), 
Johnson/Cousins ($UBV(RI)_c$) and ESO and SAAO near-infrared 
($JHKLM$) photometric systems.  Most 
of the observed objects are Herbig Ae/Be (HAeBe) stars or 
HAeBe candidates appearing in the list of HAeBe candidates of 
Th\'e et al. (1994), although several B[e] stars, LBVs and T Tauri 
are also included in our sample.  For many of the stars the data 
presented here are the first photo-electric measurements in the 
literature.  The resulting catalogue consists of 1809 photometric 
measurements.  Optical variability was detected in 66 out of the 
116 sources that were observed more than once.  15 out of the 
50 stars observed multiple times 
in the infrared showed variability at 2.2~$\mu$m ($K$ band).
}
 
\maketitle

\section{Introduction}
One of the classic characteristics of a young star is the 
presence of photometric variability.  Several types of photometric 
variability can be recognized, amongst which are more or less periodic 
variability due to magnetic activity on the stellar surface (star spots), 
brightness variations due to a variable rate of accretion, and irregular 
large-amplitude ($>$ 0\fm5) variations due to variable circumstellar 
extinction (Bibo \& Th\'e 1991; Shevchenko et al. 1993; 
Herbst et al. 1994; van den Ancker et al. 1998; Herbst \& Shevchenko 1999).  
However, a detailed analysis of a star's photometric 
behaviour requires great amounts of data, preferably over a 
large period of time.  In practice, sufficient data can often 
only be obtained by combining data sets from the literature with 
new data.

Herbig Ae/Be (HAeBe) stars may be one of the most interesting groups 
of young stars to study since they are intrinsically bright and 
therefore can be seen over great distances, tracing galactic 
regions of star formation.  Furthermore, many are sufficiently 
bright to be observed with smaller telescopes, allowing detailed 
studies of their photometric behaviour to be made.  
A catalogue of 287 candidate members of the Herbig Ae/Be stellar 
group was published by Th\'e et al. (1994).  
For many of the objects in this catalogue little or no data is 
available and therefore a new investigation of their properties 
is in order.

In this {\em paper} we present photometry of 162 emission-line and 
shell stars, most of which appear in the list of HAeBe candidates 
by Th\'e et al.  Names, positions and classifications of the stars 
are listed in Table~1.  Section~2 gives a detailed description 
of the observations.  Although a brief discussion of the data is 
presented in Section~3, we defer a detailed analysis of these 
data to future papers in which these data can be combined with 
photometry over a much longer time span.
\begin{table*}
\caption[]{Properties of programme stars}
\tabcolsep0.11cm
\begin{tabular}{llllllcccccc}
\hline\noalign{\smallskip}
Name       & Cat. Des.    & Other          & \multicolumn{1}{c}{$\alpha$ (2000)} & \multicolumn{1}{c}{$\delta$ (2000)} & Type & \multicolumn{3}{c}{\# Observations} & $V_{\rm obs}$ & $V_{\rm lit}$ & Ref.\\
\noalign{\vspace{0.02cm}}
\cline{7-9}\noalign{\vspace{0.04cm}}
           &              &  & ~~~$^{\rm h}$~~~$^{\rm m}$~~~$^{\rm s}$ & ~~~~$^{\circ}$~~~$'$~~~$''$  &     & {\scriptsize $WULBV$} & {\scriptsize $UBVRI$} & {\scriptsize $JHKLM$} & [mag] & [mag]\\
\noalign{\smallskip}
\hline\noalign{\smallskip}
71 Cet     & HD 15004     &                & 02 24 58.4 & $-$02 46 48 & A-sh & ~2 & -- & -- & 6.34--6.65   & 6.31--6.36   & (1)\\
BU Tau     & HD 23862     & MWC 75         & 03 49 11.2 &   +24 08 12 & Be   & ~1 & -- & -- & 5.08         & 5.01--5.07   & (1)\\
V1080 Tau  & HD 283817    & StH$\alpha$ 33 & 04 40 32.6 &   +24 26 31 & HAe  & ~2 & ~3 & -- & 10.10--11.68 & 10.29--10.56 & (2)\\
AB Aur     & HD 31293     & MWC 93         & 04 55 45.8 &   +30 33 04 & HAe  & ~1 & ~2 & -- & 7.02--7.04   & 7.03--7.13   & (1)\\
           & HD 31648     & MWC 480        & 04 58 46.3 &   +29 50 37 & HAe  & ~2 & ~3 & -- & 7.68--7.72   & 7.66--7.86   & (1)\\
UX Ori     & HD 293782    & 	           & 05 04 30.0 & $-$03 47 14 & HAe  & 29 & 13 & 12 & 9.84--11.79  & 9.66--11.79  & (3)\\
V1012 Ori  & 		  & 	           & 05 11 36.4 & $-$02 22 47 & HAe  & -- & ~6 & -- & 12.07--13.12 & 12.5--16.5pg & (2)\\
S22/LMC    & HD 34664     & MWC 105        & 05 13 53.1 & $-$67 26 54 & B[e] & -- & ~3 & -- & 11.77--12.65 & \\
V1366 Ori  & HD 34282     &                & 05 16 00.5 & $-$09 48 35 & HAe  & -- & ~3 & -- & 9.81--9.94   & \\
V346 Ori   & HD 287841    & 		   & 05 24 42.8 &   +01 43 48 & HAe  & 22 & ~6 & ~3 & 10.09--10.84 & 10.17--10.38 & (3)\\
CO Ori     & BD+11 809    & 	           & 05 27 38.3 &   +11 25 39 & HFe  & -- & ~3 & -- & 11.25--11.70 & 9.81--12.52  & (4)\\
           & HD 35929     &                & 05 27 42.8 & $-$08 19 38 & HFe  & -- & ~5 & ~1 & 8.11--8.31   & \\
Par 102    & BD$-$06 1193 & StH$\alpha$ 41 & 05 29 11.4 & $-$06 08 05 & HFe  & -- & ~8 & ~3 & 10.43--10.54 & \\
           & BD+11 820    & 		   & 05 29 22.1 &   +11 50 50 & HAe? & -- & ~1 & -- & 10.84        & \\
           & HD 36112     & MWC 758        & 05 30 27.5 &   +25 19 57 & HAe  & -- & ~2 & -- & 8.26--8.27   & 8.20--8.44   & (1)\\
HK Ori     & 		  & MWC 497        & 05 31 28.0 &   +12 09 11 & HFe  & 23 & ~5 & ~3 & 11.41--11.67 & 11.39--12.10 & (4)\\
           & HD 244604    &                & 05 31 57.3 &   +11 17 41 & HAe  & -- & ~3 & -- & 8.99--9.38   & 9.18--9.71   & (1)\\
RY Ori     & 		  & 	           & 05 32 10.0 & $-$02 49 49 & HFe  & ~8 & ~4 & ~3 & 11.57--12.68 & 11.08-12.96  & (3)\\
EZ Ori     & 		  & Par 1409       & 05 34 18.6 & $-$05 04 48 & TT   & -- & ~1 & -- & 12.87        & 12.0--14.1pg & (2)\\
IX Ori     &              & Par 1552       & 05 34 40.8 & $-$05 22 43 & TT   & -- & ~4 & ~1 & 12.33--13.26 & 13.4--15.5pg & (2)\\
V372 Ori   & HD 36917     & Par 1605       & 05 34 47.0 & $-$05 34 15 & A-sh & 10 & ~3 & ~3 & 7.98--8.00   & 7.98--8.03   & (3)\\
YZ Ori     &              & Par 1648       & 05 34 54:  & $-$05 03 30:& TT   & -- & ~4 & ~1 & 13.74--14.85 & 13.9--16.6pg & (2)\\
Par 1660   & HD 36939     & 		   & 05 34 55.3 & $-$05 30 22 & HBe? & ~9 & ~3 & ~3 & 8.97--9.00   & 8.98--9.02   & (3)\\
Par 1623   & BD$-$05 1306 & 		   & 05 34 55.6 & $-$05 16 43 & A-sh & 10 & ~2 & ~2 & 10.13--10.19 & \\
KS Ori     & 		  & Par 1685       & 05 35 00.1 & $-$05 25 16 & A-sh & 10 & ~2 & ~1 & 10.05--10.12 & 9.7--12.0pg  & (2)\\
V1271 Ori  & HD 245185    & 	           & 05 35 09.6 &   +10 01 52 & HAe  & -- & ~6 & -- & 9.91--9.98   & 9.85--9.95   & (3)\\
MR Ori     & 		  & 	           & 05 35 17.0 & $-$05 21 46 & A-sh & -- & ~3 & -- & 9.54--10.85  & 10.3--12.0   & (2)\\
LZ Ori     & HD 294263    & Par 1854       & 05 35 17.8 & $-$04 41 07 & A-sh & 20 & ~1 & ~3 & 9.73--10.09  & 10.3--11.1pg & (2)\\
MX Ori     & BD$-$05 1317 & Par 1953       & 05 35 21.3 & $-$05 09 16 & HFe  & -- & ~1 & -- & 9.31         & 10.3--11.7pg & (2)\\
NV Ori     & BD$-$05 1324 & Par 2086       & 05 35 31.3 & $-$05 33 09 & HFe  & ~6 & -- & ~2 & 9.63--9.90   & 9.78--10.35  & (3)\\
V361 Ori   & HD 37062     & Par 2085       & 05 35 31.4 & $-$05 25 16 & B-sh & ~9 & ~1 & ~3 & 8.15--8.20   & 8.14--8.24   & (3)\\
T Ori      & BD$-$05 1329 & MWC 763        & 05 35 50.4 & $-$05 28 35 & HAe  & 25 & 10 & ~7 & 10.15--11.04 & 10.00--11.95 & (3)\\
CQ Tau     & HD 36910     &                & 05 35 58.5 &   +24 44 54 & HFe  & ~1 & ~4 & -- & 10.22--11.06 & 9.36--11.14  & (3)\\
V380 Ori   & BD$-$06 1253 & MWC 765        & 05 36 25.4 & $-$06 42 58 & HAe  & 25 & 12 & ~7 & 10.21--10.73 & 9.46--11.22  & (4)\\
BN Ori     & HD 245465    & 	           & 05 36 29.3 &   +06 50 02 & A-sh & 11 & ~3 & ~2 & 9.62--9.68   & 9.56--9.71   & (5)\\
Par 2441   & BD$-$04 1191 & 	           & 05 36 51.2 & $-$04 25 40 & TT   & -- & ~1 & -- & 10.74        & \\
V586 Ori   & HD 37258     & Par 2473       & 05 36 59.3 & $-$06 09 16 & HAe  & 24 & ~8 & ~3 & 9.64--10.08  & 9.53--10.77  & (3)\\
BF Ori     & BD$-$06 1259 & 	           & 05 37 13.3 & $-$06 35 01 & HAe  & 25 & 14 & ~9 & 9.82--11.67  & 9.59--11.92  & (3)\\
Par 2599   & HD 37357     & KMS 27         & 05 37 47.1 & $-$06 42 30 & HAe  & -- & ~7 & ~2 & 8.83--8.84   & 8.83--8.91   & (4)\\
N3Sk81     & 		  & KMS 34         & 05 38 08.8 & $-$06 49 13 & HAe  & -- & ~6 & ~1 & 13.75--13.84 & \\
Par 2653   & HD 37411     & 		   & 05 38 14.5 & $-$05 25 13 & HAe  & ~2 & 10 & ~1 & 9.76--9.85   & 9.79--9.81   & (4)\\
           & CD$-$59 1105 & 		   & 05 38 33.0 & $-$59 04 28 & Fe   & -- & ~1 & -- & 9.74         & \\
N3Sk83     & =V883 Ori?   & KMS 40         & 05 38 18.1 & $-$07 02 26 & HBe  & -- & ~1 & ~1 & 14.4         & \\
V599 Ori   &              & KMS 49         & 05 38 54.4 & $-$07 16 38 & HAe  & -- & ~2 & ~1 & 13.76--13.80 & 15.1--16.3pg & (2)\\
$\omega$ Ori& HD 37490    & MWC 117        & 05 39 11.1 &   +04 07 17 & Be   & -- & ~3 & -- & 4.48--4.56   & 4.49--4.55   & (1)\\
V350 Ori   & 		  & 	           & 05 40 11.6 & $-$09 42 10 & HAe  & ~4 & 16 & ~3 & 10.77--13.32 & 10.71--13.24 & (3)\\
KMS 82     &              &                & 05 40 37.2 & $-$08 04 02:& HFe? & -- & ~4 & ~2 & 11.55--14.57 & \\
KMS 108    &              &                & 05 40 46.3 & $-$08 53 48:& HBe? & -- & ~2 & ~1 & 11.19--11.21 & \\
           & HD 37806     & MWC 120        & 05 41 02.3 & $-$02 43 01 & HAe  & ~9 & ~7 & ~2 & 7.86--7.96   & 7.89--7.98   & (3)\\
KMS 118    &              &                & 05 41 03.9 & $-$09 23 18:& HFe  & -- & ~3 & ~1 & 14.80--14.92 & \\
V351 Ori   & HD 38238     & 		   & 05 44 18.8 &   +00 08 40 & HAe  & ~8 & ~2 & ~4 & 8.89--9.13   & 8.81--10.94  & (6)\\
FU Ori     & BD+09 5427   & CDS 535        & 05 45 22.4 &   +09 04 12 & FUOR & ~2 & ~1 & -- & 9.09--9.32   & 8.02--10.06  & (4)\\
           & HD 288313    & 	           & 05 54 03.0 &   +01 40 22 & TT   & -- & ~7 & -- & 9.45--9.95   & \\
V1307 Ori  & HD 250550    & MWC 789        & 06 02 00.0 &   +16 30 57 & HBe  & 14 & -- & ~2 & 9.53--9.56   & 9.54--9.88   & (3)\\
17 Lep     & HD 41511     & MWC 519        & 06 04 59.1 & $-$16 29 04 & A-sh & ~2 & ~3 & -- & 4.91--5.01   & 4.97--5.02   & (1)\\
LkH$\alpha$ 208 &         &                & 06 07 49.4 &   +18 39 27 & HAe  & -- & ~3 & -- & 11.04--11.50 & 11.36--12.21 & (4)\\
LkH$\alpha$ 339 &         &                & 06 10 54.4 & $-$06 14 39 & HAe  & -- & ~5 & ~1 & 13.44--13.64 & 13.65--13.67 & (4)\\
\noalign{\smallskip}
\hline
\end{tabular}
\end{table*}
\setcounter{table}{0}
\begin{table*}
\caption[]{(Continued)}
\tabcolsep0.09cm
\begin{tabular}{llllllcccccc}
\hline\noalign{\smallskip}
Name       & Cat. Des.    & Other          & \multicolumn{1}{c}{$\alpha$ (2000)} & \multicolumn{1}{c}{$\delta$ (2000)} & Type & \multicolumn{3}{c}{\# Observations} & $V_{\rm obs}$ & $V_{\rm lit}$ & Ref.\\
\noalign{\vspace{0.02cm}}
\cline{7-9}\noalign{\vspace{0.04cm}}
           &              &  & ~~~$^{\rm h}$~~~$^{\rm m}$~~~$^{\rm s}$ & ~~~~$^{\circ}$~~~$'$~~~$''$  &     & {\scriptsize $WULBV$} & {\scriptsize $UBVRI$} & {\scriptsize $JHKLM$} & [mag] & [mag]\\
\noalign{\smallskip}
\hline\noalign{\smallskip}
V1308 Ori  &              & MWC 137        & 06 18 45.5 &   +15 16 52 & HBe  & -- & ~2 & -- & 11.67--11.98 & 11.79--12.27 & (4)\\
FS CMa     & HD 45677     & MWC 142        & 06 28 17.4 & $-$13 03 11 & B[e] & 17 & 13 & ~2 & 7.94--8.74   & 7.22--8.85   & (7)\\
VY Mon     & 		  & 	           & 06 31 06.9 &   +10 26 05 & HBe  & -- & ~3 & -- & 12.97--13.09 & 12.77--14.77 & (4)\\
V699 Mon   & 		  & LkH$\alpha$ 215& 06 32 41.8 &   +10 09 34 & HBe  & 19 & ~5 & ~2 & 10.46--10.91 & 10.36--10.84 & (2)\\
V700 Mon   & HD 259431    & MWC 147        & 06 33 05.2 &   +10 19 20 & HBe  & 13 & ~6 & ~2 & 8.69--8.75   & 8.65--8.97	  & (1)\\
R Mon      & BD+08 1427   & MWC 151        & 06 39 09.9 &   +08 44 10 & HBe  & 13 & ~4 & ~2 & 10.35--12.41 & 11.60--12.54 & (4)\\
V590 Mon   & W90          & LkH$\alpha$ 25 & 06 40 44.6 &   +09 48 02 & HFe  & -- & ~8 & ~1 & 11.88--12.94 & 12.60--12.95 & (4)\\
ST Pup     & CD$-$37 3101 &                & 06 48 56.4 & $-$37 16 33 & Ceph.& -- & ~1 & -- & 10.31        & 9.28--10.68  & (2)\\
V743 Mon   & HD 50138     & MWC 158        & 06 51 33.4 & $-$06 57 59 & B[e] & 20 & 13 & ~2 & 6.59--6.85   & 6.53--6.65   & (1)\\
GU CMa     & HD 52721     & MWC 164        & 07 01 49.5 & $-$11 18 03 & HBe  & ~1 & ~2 & -- & 6.50--6.57   & 6.52--6.71   & (1)\\
Z CMa      & HD 53179     & MWC 165        & 07 03 43.2 & $-$11 33 06 & FUOR & 19 & 19 & ~6 & 8.68--10.27  & 8.61--10.22  & (4)\\
HU CMa     &              & LkH$\alpha$ 220& 07 04 06.8 & $-$11 26 07 & HBe  & -- & ~3 & -- & 11.52--11.92 & 11.57--12.19 & (4)\\
V750 Mon   & HD 53367     & MWC 166        & 07 04 25.5 & $-$10 27 16 & HBe  & 12 & ~4 & ~3 & 6.92--7.08   & 7.00--7.22   & (1)\\
NX Pup     & CD$-$44 3318 & Hen 3-32       & 07 19 28.3 & $-$44 35 11 & HAe  & 15 & ~7 & ~3 & 9.63--11.08  & 9.53--10.99  & (3)\\
V694 Mon   & 		  & MWC 560        & 07 25 51.3 & $-$07 44 08 & Symb.& -- & ~8 & -- & 9.96--10.33  & 9.10--10.10  & (2)\\
           & HD 58647     &                & 07 25 56.1 & $-$14 10 44 & A-sh & -- & -- & ~1 &              & 6.76--6.88   & (1)\\
           & HD 59319     & MWC 843        & 07 28 36.8 & $-$21 57 49 & HBe? & -- & ~4 & -- & 8.29--8.31   & 8.19--8.47   & (1)\\
           & HD 59771     & 		   & 07 30 51.1 & $-$18 15 43 & Fe   & -- & ~2 & -- & 8.89--8.94   & 8.83--9.31   & (1)\\
Hen 3-40   &              & Wray 15-30     & 07 31 48:  & $-$41 34 00:& Ge   & -- & ~3 & ~1 & 13.73--14.42 & \\
PW Pup     & CD$-$30 5135 & Hen 3-83       & 07 49 06.0 & $-$31 07 43 & Fe   & -- & ~2 & -- & 9.23--9.24   & 9.27--9.97   & (1)\\
V402 Pup   & HD 64315     & 		   & 07 52 20.3 & $-$26 25 47 & Oe   & -- & ~1 & -- & 9.24         & 8.99--9.39   & (1)\\
AS 202     & CD$-$37 4833 & Hen 3-174      & 08 32 35.8 & $-$37 59 02 & HBe? & -- & ~1 & -- & 10.80        & 9.70--10.70  & (2)\\
ESO 313-10 &              &                & 08 42 17.0 & $-$40 44 13:& HBe  & -- & ~3 & ~1 & 14.24--14.59 & \\
Hen 3-209  & 		  & Wra 15-285     & 08 48 45.5 & $-$46 05 08 & B[e] & -- & ~1 & -- & 13.56        & \\
Hen 2-14   & 		  & Wra 16-30      & 08 51 59.5 & $-$46 18 05:& B[e] & -- & ~1 & -- & 14.05        & \\
OU Vel     & HD 76534     & Hen 3-225      & 08 55 08.7 & $-$43 28 00 & HBe  & ~2 & ~2 & ~2 & 7.96--8.08   & \\
RCW 34     &              & Gum 19         & 08 56 27.8 & $-$43 05 46:& O    & -- & ~2 & -- & 11.21--11.64 & \\
Herbst 28  &              & Bran 215       & 08 58 26.3 & $-$43 26 10 & Be   & -- & ~2 & ~1 & 11.25--14.76 & \\
RCW 36     &              & Bran 217       & 08 59 00.9 & $-$43 44 10 & HFe? & -- & ~1 & -- & 15.20        & \\
           & HD 309784    & Hen 3-315      & 09 42 36.4 & $-$67 08 54 & B[e] & -- & ~1 & -- & 10.16        & \\
           & HD 85567     & Hen 3-331      & 09 50 28.5 & $-$60 58 03 & HBe  & -- & 33 & -- & 8.49--8.59   & 8.44--8.75   & (1)\\
           & HD 87643     & MWC 198        & 10 04 30.3 & $-$58 39 52 & B[e] & 22 & 16 & ~3 & 8.52--9.00   & 8.68--9.20   & (3)\\
17 Sex     & HD 88195     &                & 10 10 07.5 & $-$08 24 29 & B-sh & ~1 & -- & -- & 5.90         & 5.87--5.93   & (1)\\
           & HD 89249     & MWC 200        & 10 16 20.6 & $-$55 35 51 & Be   & -- & ~1 & ~1 & 8.90         & 8.70--9.15   & (1)\\
HR Car     & HD 90177     & MWC 202        & 10 22 53.8 & $-$59 37 28 & LBV  & 21 & 11 & ~3 & 7.44--8.42   & 7.40--8.34   & (3)\\
Hen 3-416  & 		  & Th 35-40       & 10 25 44.5 & $-$58 33 52 & HBe? & -- & ~1 & -- & 10.59        & \\
Wra 15-566 &              & 		   & 10 25 51.7 & $-$60 53 14 & TT?  & -- & ~2 & -- & 13.87--13.90 & \\
V503 Car   & HD 90578     & Hen 3-418      & 10 26 00.2 & $-$57 49 37 & B[e] & -- & ~1 & -- & 9.33         & 9.08--9.75   & (1)\\
           & HD 92061     & Hen 3-455      & 10 36 24.2 & $-$58 57 09 & HBe? & -- & ~1 & -- & 8.97         & 8.99--9.87   & (1)\\
           & CPD$-$59 2617& Hen 3-480      & 10 44 49.4 & $-$59 49 28:& HBe? & -- & -- & ~1 &              & \\
           & HD 94509     & Hen 3-515      & 10 53 27.3 & $-$58 25 25 & HBe  & -- & ~4 & -- & 9.08--9.13   & 8.85--9.41   & (1)\\
           & CPD$-$59 2854& Wra 15-689     & 10 55 55.4 & $-$60 14 20 & B[e] & -- & ~1 & -- & 10.54        & \\
GG Car     & HD 94878     & MWC 215        & 10 55 58.9 & $-$60 23 33 & B[e] & 23 & 14 & ~3 & 8.53--9.03   & 8.48--9.00   & (8)\\
           & HD 305773    & CD$-$59 3426   & 10 56 03.9 & $-$60 29 38 & HBe? & -- & ~1 & -- & 9.08         & 8.80--9.39   & (1)\\
AG Car     & HD 94910     & MWC 216        & 10 56 11.6 & $-$60 27 13 & LBV  & 22 & 11 & ~3 & 6.49--7.00   & 6.10--8.12   & (9)\\
HR 4329    & HD 96706     &                & 11 06 49.9 & $-$70 52 41 & B-sh & ~2 & -- & -- & 5.58--5.59   & 5.54--5.60   & (1)\\
DI Cha     & CD$-$76 486  & Hen 3-593      & 11 07 20.7 & $-$77 38 07 & TT   & 23 & ~3 & ~3 & 10.65--10.74 & 10.66--10.74 & (10)\\
CU Cha     & HD 97048     & Hen 3-597      & 11 08 03.3 & $-$77 39 17 & HAe  & 23 & ~9 & ~7 & 8.40--8.48   & 8.44--8.48   & (10)\\
           & HD 97240     &                & 11 09 18.1 & $-$77 47 40 & F-sh & ~2 & -- & -- & 8.49--8.50   & 8.42--8.69   & (1)\\
           & HD 97300     &                & 11 09 50.0 & $-$76 36 48 & B-sh & -- & ~4 & ~1 & 9.02--9.08   & 8.97--9.03   & (1)\\
           & HD 98922     & Hen 3-644      & 11 22 31.7 & $-$53 22 11 & HBe  & ~2 & 38 & -- & 6.72--6.81   & 6.72--6.83   & (1)\\
KR Mus     & HD 100546    & Hen 3-672      & 11 33 25.4 & $-$70 11 41 & HBe  & ~2 & ~4 & ~1 & 6.69--6.74   & 6.71--6.77   & (3)\\
           & HD 101412    & Hen 3-692      & 11 39 44.5 & $-$60 10 28 & HBe  & -- & ~6 & ~1 & 9.24--9.28   & 8.97--9.56   & (1)\\
V644 Cen   & HD 306989    & Hen 3-700      & 11 43 06.5 & $-$60 44 05 & Be   & -- & ~1 & ~1 & 10.52        & 9.91--10.33  & (10)\\
DX Cha     & HD 104237    & Hen 3-741      & 12 00 05.1 & $-$78 11 35 & HAe  & ~2 & 38 & ~1 & 6.49--6.59   & 6.56--6.68   & (1)\\
           & HD 105234    &                & 12 07 05.5 & $-$78 44 28 & A-sh & ~2 & -- & -- & 7.48--7.48   & 7.45--7.59   & (1)\\
RU Cen     & HD 105578    & Hen 3-755      & 12 09 23.8 & $-$45 25 35 & Ge   & -- & ~1 & -- & 8.63         & 8.78--9.56   & (1)\\
\noalign{\smallskip}
\hline
\end{tabular}
\end{table*}
\setcounter{table}{0}
\begin{table*}
\begin{flushleft}
\caption[]{(Continued)}
\tabcolsep0.08cm
\begin{tabular}{llllllcccccc}
\hline\noalign{\smallskip}
Name       & Cat. Des.    & Other          & \multicolumn{1}{c}{$\alpha$ (2000)} & \multicolumn{1}{c}{$\delta$ (2000)} & Type & \multicolumn{3}{c}{\# Observations} & $V_{\rm obs}$ & $V_{\rm lit}$ & Ref.\\
\noalign{\vspace{0.02cm}}
\cline{7-9}\noalign{\vspace{0.04cm}}
           &              &  & ~~~$^{\rm h}$~~~$^{\rm m}$~~~$^{\rm s}$ & ~~~~$^{\circ}$~~~$'$~~~$''$  &     & {\scriptsize $WULBV$} & {\scriptsize $UBVRI$} & {\scriptsize $JHKLM$} & [mag] & [mag]\\
\noalign{\smallskip}
\hline\noalign{\smallskip}
Hen 3-759  &              &                & 12 12 08.6 & $-$62 29 01 & Be   & -- & ~1 & -- & 10.47        & \\
DK Cha     & IRAS12496$-$7650 &            & 12 53 16.1 & $-$77 07 02:& HAe  & -- & -- & ~1 &              & \\
V1028 Cen  & CD$-$48 7859 & Hen 3-847      & 13 01 17.8 & $-$48 53 19 & HAe  & -- & 34 & ~2 & 10.56--10.65 & \\
LSS 3027B  &CPD$-$61 3587B& 		   & 13 19 04.0 & $-$62 34 10 & HBe  & ~1 & ~2 & ~3 & 15.17--15.24 & \\
CQ Cir     & HD 130437    & Hen 3-1031     & 14 50 50.3 & $-$60 17 10 & HBe  & -- & ~7 & ~1 & 9.85--9.91   & \\
SS73 44    &              &                & 15 03 23.9 & $-$63 22 54 & HBe  & -- & ~2 & ~1 & 14.75--16.80 & \\
           & HD 132947    &                & 15 04 56.1 & $-$63 07 53 & HAe  & -- & ~1 & -- & 8.91         & 8.73--9.12   & (1)\\
HT Lup     & CD$-$33 10685& Hen 3-1095     & 15 45 12.9 & $-$34 17 31 & TT   & ~9 & -- & -- & 10.21--10.26 & 10.26--10.40 & (2)\\
           & HD 140817    &                & 15 47 04.5 & $-$35 30 37 & B-sh & ~2 & -- & -- & 6.83--6.84   & 6.77--6.90   & (1)\\
           & HD 141569    & 		   & 15 49 57.7 & $-$03 55 16 & HAe  & -- & ~1 & -- & 7.10         & 7.11--7.14   & (3)\\
           & HD 144432    & Hen 3-1141     & 16 06 58.0 & $-$27 43 10 & HAe  & -- & ~1 & -- & 8.17         & 8.08--8.35   & (1)\\
V856 Sco   & HD 144668    & HR 5999        & 16 08 34.3 & $-$39 06 18 & HAe  & ~9 & 26 & 13 & 6.55--7.32   & 6.64--7.95   & (3)\\
V1027 Sco  & HD 144667    & HR 6000        & 16 08 34.6 & $-$39 05 34 & B-sh & ~7 & ~8 & ~9 & 6.65--6.73   & 6.65--6.67   & (3)\\
           & HD 147196    & 		   & 16 21 19.2 & $-$23 42 29 & Be   & -- & ~5 & ~2 & 7.00--7.04   & 7.02--7.12   & (1)\\
Wra 15-1484&              & Hen 3-1191     & 16 27 14.2 & $-$48 39 28 & B[e] & -- & 44 & -- & 12.52--13.91 & \\
V2307 Oph  & HD 150193    & MWC 863        & 16 40 17.9 & $-$23 53 45 & HAe  & ~3 & 11 & ~1 & 8.80--8.88   & 8.79--8.87   & (10)\\
AK Sco     & HD 152404    &                & 16 54 44.8 & $-$36 53 19 & HFe  & ~3 & ~5 & ~2 & 8.81--9.18   & 8.76--9.93   & (3)\\
AS 215     & HD 322422    & Hen 3-1293     & 16 57 23.9 & $-$40 21 40 & B[e] & -- & ~6 & -- & 9.68--9.71   & 9.71--9.75   & (3)\\
V921 Sco   & CD$-$42 11721& MWC 865        & 16 59 06.9 & $-$42 42 08 & B[e] & ~4 & ~7 & ~4 & 11.10--11.36 & 11.43--11.73 & (2)\\
V1104 Sco  & HD 326823    & Hen 3-1330     & 17 06 53.9 & $-$42 36 40 & B[e] & -- & ~8 & -- & 8.98--9.04   & 8.90--9.33   & (1)\\
KK Oph     & 	          & Hen 3-1346     & 17 10 08.1 & $-$27 15 18 & HAe  & ~3 & ~7 & ~1 & 11.32--12.36 & 11.37--11.87 & (3)\\
Wra 15-1651&              & 		   & 17 14 45.6 & $-$36 18 35:& HBe? & -- & -- & ~1 &              & \\
Wra 15-1678&              & 		   & 17 20 14.8 & $-$37 39 35:& HBe? & -- & -- & ~1 &              & \\
           & HD 156702    & MWC 257        & 17 20 50.6 & $-$38 39 09 & B[e] & -- & ~4 & -- & 8.70--8.72   & 8.58--8.83   & (1)\\
51 Oph     & HD 158643    & HR 6519        & 17 31 25.0 & $-$23 57 46 & Be   & ~2 & ~7 & -- & 4.77--4.81   & 4.77--4.82   & (1)\\
XX Oph     & HD 161114    & MWC 269        & 17 43 56.5 & $-$06 16 09 & Symb.& 11 & ~5 & ~1 & 8.80--9.08   & 8.80--9.11   & (3)\\
           & HD 163296    & MWC 275        & 17 56 21.3 & $-$21 57 22 & HAe  & ~8 & 15 & ~4 & 6.82--6.93   & 6.82--6.93   & (1)\\
LkH$\alpha$ 108 &         &                & 18 03 50.6 & $-$24 21 11 & HBe? & ~2 & -- & -- & 11.67--11.71 & \\
V4203 Sgr  & CD$-$24 13830& LkH$\alpha$ 112& 18 04 22.5 & $-$24 22 10 & HBe  & ~2 & ~2 & ~1 & 9.95--9.98   & 9.54--10.19  & (4)\\
           & HD 164906    & MWC 280        & 18 04 25.8 & $-$24 23 08 & Oe   & ~3 & ~2 & -- & 7.46--7.49   & 7.32--7.59   & (1)\\
LkH$\alpha$ 115 &         &                & 18 04 50.4 & $-$24 25 42 & HBe? & ~3 & ~2 & ~1 & 11.95--11.96 & 10.80--12.25 & (4)\\
LkH$\alpha$ 118 & CD$-$24 13874& Hen 3-1583& 18 05 49.6 & $-$24 15 20 & Be   & ~5 & ~6 & ~3 & 11.11--11.18 & 10.69--11.61 & (4)\\
LkH$\alpha$ 119 &         &                & 18 05 56:  & $-$24 14 24:& Be   & ~5 & ~6 & ~3 & 12.15--12.21 & 11.27--12.76 & (4)\\
NZ Ser     & 		  & MWC 297        & 18 27 39.6 & $-$03 49 52 & HBe  & ~6 & ~5 & -- & 12.03--12.63 & 11.95--12.60 & (4)\\
VV Ser     & 		  & 	           & 18 28 47.9 &   +00 08 40 & HAe  & ~1 & ~4 & -- & 11.81--12.31 & 11.18--13.09 & (4)\\
V431 Sct   & 		  & MWC 300        & 18 29 25.7 & $-$06 04 37 & B[e] & ~1 & ~1 & -- & 11.73--11.79 & 11.54--11.90 & (4)\\
AS 310     & 	         &MH$\alpha$ 375-17& 18 33 21.2 & $-$04 58 07 & HBe  & ~1 & ~1 & ~4 & 12.47--12.59 & 12.08--13.19 & (4)\\
HR 7169    & HD 176269    &                & 19 01 03.2 & $-$37 03 39 & B-sh & ~2 & -- & -- & 6.72--6.74   & 6.63--6.73   & (1)\\
HR 7170    & HD 176270    &                & 19 01 04.3 & $-$37 03 42 & B-sh & ~1 & -- & -- & 6.45         & 6.34--6.44   & (1)\\
S CrA      &              & Hen 3-1731     & 19 01 20:  & $-$36 57 00:& TT   & ~8 & -- & -- & 10.46--11.76 & 10.80--12.44 & (4)\\
           & HD 176386    &                & 19 01 38.9 & $-$36 53 27 & B-sh & 11 & ~6 & ~3 & 7.27--7.43   & 7.27--7.30   & (3)\\
TY CrA     & CD$-$37 13024& 	           & 19 01 40.8 & $-$36 52 34 & A-sh & 12 & ~6 & ~5 & 9.34--9.70   & 9.06--9.79   & (3)\\
R CrA      & CD$-$37 13027& Hen 3-1733     & 19 01 53.7 & $-$36 57 08 & HAe  & ~8 & 10 & ~6 & 12.10--13.72 & 10.18--14.65 & (4)\\
T CrA      &              &                & 19 01 58.8 & $-$36 57 49 & HFe  & ~7 & ~2 & ~4 & 13.27--14.10 & 12.10--13.99 & (4)\\
           & HD 179218    & MWC 614        & 19 11 11.3 &   +15 47 16 & HBe  & ~1 & -- & -- & 7.39         & 7.34--7.47   & (1)\\
WW Vul     & HD 344361    & MWC 987        & 19 25 58.8 &   +21 12 31 & HAe  & 10 & ~8 & -- & 10.48--11.68 & 10.25--12.40 & (4)\\
V1295 Aql  & HD 190073    & MWC 325        & 20 03 02.5 &   +05 44 17 & HAe  & ~1 & -- & -- & 7.86         & 7.84--7.89   & (10)\\
HR 7836    & HD 195325    & MWC 1019       & 20 30 18.0 &   +10 53 45 & HBe? & ~2 & -- & -- & 5.97--6.03   & 6.00--6.09   & (1)\\
\noalign{\smallskip}
\hline\noalign{\smallskip}
Total:     & 162 stars    &                &            &             &     & 753~&830~&229~& \\
\noalign{\smallskip}
\hline
\end{tabular} \\
\vspace*{0.1cm}
\noindent\small
Explanation of the abbreviations used: HAe/Be = Herbig Ae/Be star; 
HFe = F-type Herbig-like star; TT = T Tauri star; FUOR = FU Orionis type 
star; A-, B-sh: A- or B- type star with shell lines; Be = classical 
Be star; B[e] = B[e] star; Symb. = symbiotic star; Ceph. = Cepheid; 
LBV = Luminous Blue Variable.\\
\smallskip

References:
(1) Tycho Catalogue (ESA 1997); 
(2) General Catalogue of Variable Stars (Kholopov et al. 1998); 
(3) ESO Long Term Photometry of Variables data (Manfroid et al. 
 1991, 1995; Sterken et al. 1993, 1995); 
(4) Herbst \& Shevchenko (1999);
(5) Shevchenko et al. (1997); 
(6) van den Ancker et al. (1996); 
(7) de Winter \& van den Ancker (1997); 
(8) Gosset et al. (1984);
(9) Sterken et al. (1996); 
(10) Kilkenny et al. (1985).

\end{flushleft}
\end{table*}

\section{Observations}
Photometric data in the Walraven $WULBV$ system of our programme 
stars were obtained between 1978 and 1988 at the 90 cm Dutch Light 
Collector.  The data taken in 1978 were obtained when this 
telescope was located in Hartebeespoortdam, South Africa.  All 
observations in later years were obtained after the telescope had 
been moved to the European Southern Observatory, La Silla, Chile. 
A detailed description of the Walraven photometric system can be 
found in de Geus et al. (1989), who also explain the employed 
measuring and data reduction procedures.  All data presented 
here were made with a 5-channel photometer which 
measures all Walraven $WULBV$ intensities simultaneously, 
thereby avoiding any systematic effects due to non-simultaneous
measurements at the different photometric passbands.
A circular diaphragm with a diameter of 21\farcs5 was used 
for all stars.  Typical errors in the data, given in 
$\log_{10}$ intensity values, are 0.004, 0.003, 0.004, 0.006, 
and 0.009 for $V$, $V-B$, $B-U$, $U-W$, and $B-L$, respectively. 
Results of the photometric measurements in the Walraven system 
are listed in Table~2 (only available electronically).  The 
last column in this table lists the corresponding $V$ magnitude in 
the Johnson photometric system, computed using the transformation 
formula by Brand \& Wouterloot (1988).  Also listed in Table~2 
are the date and heliocentric julian date (JD) of each observation.  
In some cases, the exact universal time of individual observations 
proved impossible to reconstruct.  In those cases, only the integer 
fraction of the JD is listed.

Optical photometry in the Johnson/Cousins $UBV(RI)_c$ system, was obtained 
at La Silla with the ESO 50 cm and 1 m telescopes, during several observing 
runs between 1979 and 1994.  Additional $UBV(RI)_c$ photometry was obtained 
in February 1997 at the South African Astronomical Observatory (SAAO) using 
the SAAO 75 cm telescope.  During these observations, mostly made through 
a 15\arcsec\ circular diaphragm, the telescope was equipped with a single 
channel photometer and either a Quantacon RCA 3103A (pre-1993) or EMI 9658RA 
(post-1993) photo-multiplier.  The observations through $U$, $B$, $V$, $R_c$ 
and $I_c$ filters were made consecutively, in ascending or in descending order.  
Data were reduced with the Dutch method, using the E-region 
standard stars from the list by Graham (1982).  Non-variable K and M0 
standards were always included to avoid large transformation errors.  
Typical errors in the resulting data, listed in Table~3 (only available 
electronically), are 0\fm01, 0\fm01, 0\fm02, 0\fm01, 0\fm02 for $V$, $B-V$, 
$U-B$, $V-R$ and $V-I$, respectively.  For stars fainter than 12$^{\rm th}$ 
magnitude the errors may be twice those values.  A colon following the 
data indicates that these particular data are more uncertain than the values 
listed here.  Two colons indicates very uncertain data.

Near-IR photometric data in the ESO $JHKLM$ system (Bouchet et al. 1989, 1991), 
of the programme stars were obtained with the ESO 1 m telescope at La Silla 
on several occasions between 1979 and 1995.  Additional Near-IR observations  
in the SAAO $JHKLM$ system (Carter 1990; Carter \& Meadows 1995) were obtained 
with the 1.9 m telescope at the SAAO in 1997.  During the near-IR observations, 
made through a 13\arcsec\ diaphragm, the telescope was equipped with a 
photovoltaic InSb detector unit.  Standard stars were obtained from the list 
by Bouchet et al. (1991).  
Sky subtraction was achieved by chopping, with a frequency of 8 Hz, in 
the east-west direction with a throw of 30\arcsec\ amplitude.  The $J$, $H$, 
$K$, $L$ and $M$ magnitudes were again measured consecutively, in ascending 
or in descending order.  Typical errors in 
these data are 0\fm04, 0\fm04, 0\fm03, 0\fm05, 0\fm06 for the $J$, $H$, $K$, $L$ 
and $M$ magnitudes, respectively.  The resulting near-IR magnitudes of our 
programme stars are listed in Table~4 (only available electronically).

\section{Discussion}
A summary of all 1809 new photometric measurements is given in the 
last four columns of Table~1, which list the number of obtained 
measurements in the Walraven, Johnson/Cousins and Near-IR photometric 
systems, as well as the observed range in $V$ magnitude.  
Note that for some stars with a small range in $V$, this is 
entirely compatible with the expected errors in our measurements.  
Also listed in Table~1 is the range of photometric values found 
in the literature.  Ranges derived from photographic rather than 
photo-electric data are indicated by the suffix ``pg''.  In most 
cases the range observed by us is within the previously known 
range.  In other cases (e.g. V1080 Tau), the data presented here 
expands the known variability range.  In some objects (e.g. 
LkH$\alpha$ 339, V921 Sco) the range in $V$ magnitude observed 
by us is completely out of the range known from literature.  In 
those cases one might expect the star to show variations on time 
scales longer than those covered by our data.

Although a detailed analysis of all the data presented here is 
beyond the scope of this paper, we note that the statistical 
properties of the observed variations are in good agreement 
with previous work.  Of the 116 stars with more than one measurement 
in Tables 2 or 3, 66 (or 57\%) show variability in the $V$ band 
with a magnitude exceeding 0\fm1.  This is very close to the 
65\% of HAeBes showing significant variations in 
the study by van den Ancker et al. (1998).  The slightly 
lower fraction found here can easily be explained by the 
smaller number of measurements per star in the current 
study, causing us to not detect variability in some 
known variables.

In the case of near-infrared variability, literature data is 
much more scarce (in fact, for many of the stars the data 
listed in Table~4 are the first $JHKLM$ measurements 
in the literature).  Therefore a comparison with previous 
work is more cumbersome.  
Of the 50 stars with more than one measurement in Table 4, 15 
show significant ($>$ 0\fm2) variability in their $K$ (2.2~$\mu$m) 
magnitude.  Again we must caution that in view of the limited 
amount of measurements per star this will certainly be an 
underestimate of the number of near-infrared variables in our 
sample.  In fact, in some cases the magnitudes we provide 
differ by more than 0\fm5 with determinations from the literature 
(Gezari et al. 1999; Eiroa et al. 2001).  This could indicate 
variability of the circumstellar disk around these objects, 
rather than the variable circumstellar obscuration commonly 
associated with large-amplitude optical brightness variations.  
Consequently these stars are good candidates for further studies.

The data presented here are suitable to be combined with the 
photometric data of the studied stars already present in the literature 
to perform a more detailed analysis of the photometric behaviour
of the stars in our sample.  However, the main value of the 
catalogue presented here may be for historical reference.  
It is well known that some young stars with well-documented historical 
variability (e.g. AB Aur, BN Ori, V351 Ori) have in recent years 
displayed a constant brightness for decades or longer.  It is likely 
that some stars in our sample which have remained relatively constant 
over the time-frame studied here will become more active in years to 
come and vice-versa, that our variable stars may at some time cease 
to vary.  In those cases, the information presented in Tables 2--4 
will be extremely valuable in exploring the presently unknown 
cause of such long-term changes in photometric behaviour of 
young stars.

\acknowledgements{The authors would like to thank the staff at ESO La Silla 
and at the SAAO for the excellent help during the various observing runs.  
We would also like to thank several persons who spent long nights observing 
some of the data listed in this paper, and were not mentioned before.  
In alphabetic order: S. van Amerongen, E.A. Bibo, H. Cuypers, A. Hollander, 
M. Janssens, Y.K. Ng, R. van Ojik, L. Remijn and J.M. Smit. 
We are also indebted to Dr. A. van Genderen for his many years of 
work into maintaining the Walraven photometry programme on the Dutch 
telescope at La Silla.  The referee, C. Catala, provided many suggestions 
for improvements of the paper, for which we thank him.  This research has 
made use of the Simbad data base, operated at CDS, Strasbourg, France.
}


\onecolumn

\clearpage


\clearpage
\begin{verbatim}
Table 4.  Photometry of programme stars in the ESO/SAAO JHKLM system
----------------------------------------------------------------------------------
Star        Date       JD-2400000    J       H       K       L       M    Remarks
----------------------------------------------------------------------------------
UX Ori      09-11-79   44187        8.93    8.61    7.64    6.61    --
            09-11-79   44187        9.03    8.57    7.66    6.58    --
            09-11-79   44187        8.91    8.57    7.62    6.63    --
            10-11-79   44188        9.46    8.55    7.79    6.50    5.75:
            20-09-80   44503        9.26    8.30    7.36    6.17    --
            26-09-80   44509        9.12    8.28    7.35    6.24    --
            14-01-81   44619.528    9.36    8.41    7.40    6.19    5.82
            15-01-81   44620.566    9.32    8.40    7.38    6.16    5.43
            16-01-81   44621.563    9.06    8.31    7.36    6.10    5.51
            24-02-86   46486        9.44    9.11    9.02    8.46    --
            26-02-86   46488        8.03    7.43    6.71    5.61    5.32
            04-02-88   47196        8.87    8.14    7.29    6.02:   --
	    							     
V346 Ori    14-01-81   44619.542    9.84    9.01    8.23    7.27    6.46
            15-01-81   44620.579    9.59    8.97    8.21    7.12    6.33
            16-01-81   44621.578    9.46    8.97    8.21    7.11    6.53

HD 35929    28-01-97   50477.341    7.22    6.87    6.59    5.96    5.14

BD-06 1193  24-07-86   46636        9.91    9.18    8.52    7.29    7.34  
            26-07-86   46638        9.73    9.05    8.45    7.36    6.48
            28-01-97   50477.350    9.06    8.40    7.84    6.98    5.78:

HK Ori      14-01-81   44619.616    --      8.50    7.40    --      5.07
            15-01-81   44620.619    9.59    8.53    7.58    5.78    5.28
            16-01-81   44621.588    9.46    8.47    7.40    5.75    5.10

RY Ori      25-07-86   46637        9.91    9.18    8.52    7.29    7.34
            27-07-86   46639        9.73    9.05    8.45    7.36    6.48
            03-02-97   50483.389    9.72    8.95    8.32    7.48    --

IX Ori      28-01-97   50477.380   10.48    9.36    8.56    7.45    --

V372 Ori    14-01-81   44619.553    7.35    7.05    6.58    5.45    5.02
            15-01-81   44620.636    7.35    7.08    6.60    5.48    4.87
            16-01-81   44621.503    7.23    7.06    6.56    5.51    4.83

YZ Ori      29-01-97   50478.390    7.29    6.95    6.47    5.49    --   

HD 36939    15-01-81   44620.742    8.92    8.87    8.82    8.85    --
            16-01-81   44621.529    8.96    8.94    8.92    8.37    --
            29-01-97   50478.371    7.29    6.95    6.47    5.49    --

BD-05 1306  15-01-81   44620.749    8.51    8.00    7.45    6.41    5.70
            16-01-81   44621.521    8.59    8.05    7.51    6.40    5.82
            			 				     
KS Ori      15-01-81   44620.737    9.44    9.13    9.23    8.78    --
            			 				     
LZ Ori      14-01-81   44619.732    8.93    --      8.43    8.39    7.28
            15-01-81   44620.557    8.76    8.57    8.42    8.08    7.52
            16-01-81   44621.597    8.75    8.57    8.42    8.12    --
            			 				     
NV Ori      24-02-86   46486.264    8.94    8.38    7.73    6.60    6.52  
            26-02-86   46488.268    8.99    8.39    7.72    6.49    6.64
	    
V361 Ori    14-01-81   44619.565    8.02    7.74    7.60    7.03    6.52
            15-01-81   44620.647    7.87    7.76    7.60    6.99    6.48
            16-01-81   44621.512    7.80    7.69    7.56    6.97    6.68
            			 				     
T Ori       09-11-79   44187        8.13    7.52    6.40    5.55    --
            10-09-79   44188        8.40    7.53    6.40    5.31    4.83
            20-09-80   44503        8.43    7.35    6.31    5.05    4.53
            26-09-80   44509        8.52    7.38    6.31    5.11    4.67
            14-01-81   44619.706    8.78    7.59    6.45    5.17    4.86
            15-01-81   44620.637    8.71    7.54    6.43    5.14    4.78
            16-01-81   44621.556    8.48    7.58    6.44    5.17    4.76
            			  				     
V380 Ori    09-11-79   44187        7.85    7.31    6.22    5.05    --
            10-11-79   44188        8.31    7.78    6.44    4.90    4.63
            20-09-80   44503        8.27    7.14    6.10    4.51    3.97
            26-09-80   44509        8.20    7.13    5.98    4.50    3.86
            13-01-81   44618.688    7.92    7.05    5.99    4.53    3.94
            15-01-81   44620.683    8.12    7.08    5.96    4.57    3.97
            16-01-81   44621.572    7.97    7.12    6.02    4.54    4.06

BN Ori      23-02-86   46485        8.74    8.41    8.33    9.77    -- 
            25-02-86   46487        8.68    8.36    8.25    9.26    -- 

V586 Ori    13-01-81   44618.697    9.04    8.57    7.86    6.58    5.78
            15-01-81   44620.703    9.14    8.58    7.80    6.56    5.76
            16-01-81   44621.658    9.13    8.54    7.82    6.67    6.38:
            			 				     
BF Ori      09-11-79   44187        8.68    8.64    7.92    7.07    --
            10-11-79   44188        9.08    8.64    8.87    7.51    6.21
            20-09-80   44503        9.22    8.56    7.79    6.63    --
            26-09-80   44509        9.30    8.58    7.81    6.60    --
            13-01-81   44618.708    9.06    8.46    7.72    6.55    5.92
            14-01-81   44619.742    9.06    8.51    7.84    6.67    5.91
            15-01-81   44620.694    9.15    8.46    7.69    6.53    6.11
            25-02-86   46487        9.34    8.75    7.99    6.69    6.21
            26-02-86   46488        9.23    8.68    7.95    6.61    6.56
            							     
HD 37357    02-02-97   50482.342    8.52    8.12    7.56    6.66    6.17:
            03-02-97   50483.319    8.50    8.11    7.53    6.58    6.09:

N3Sk81      29-01-97   50478.431    9.93    8.67    7.61    6.43    5.86:   

HD 37411    01-02-97   50481.368    9.08    8.34    7.55    6.45    5.92:

N3Sk83      01-02-97   50481.387    9.64    6.87    5.16    3.44    2.50:   

V599 Ori    28-01-97   50477.420    9.81    8.59    7.60    6.44    5.80:

V350 Ori    03-02-87   46829.635    9.73    8.98    8.13    7.07    --  
            04-02-87   46830.635    9.75    8.95    8.12    6.99    --
            05-02-87   46831.635    9.76    8.97    8.15    6.83    --

KMS 82      02-02-97   50482.361   10.13    8.84    7.96    7.20    --
            03-02-97   50483.350    9.86    8.76    7.90    6.76    6.78:

KMS 108     01-01-97   50481.408   10.00    9.90    9.74    9.0::   --    

HD 37806    13-01-81   44618.721    7.22    6.55    5.67    4.63    3.77
            16-01-81   44621.674    7.33    6.64    5.71    4.39    3.90
	    			
KMS 118     03-02-97   50483.370   12.97   12.58   12.20    --      --    

V351 Ori    14-09-95   49975.400    --      7.41    6.70    5.82    5.3
            16-09-95   49977.424    7.99    7.48    6.78    5.96    -- 
            18-09-95   49979.374    8.02    7.49    6.82    5.86    -- 
            03-02-97   50483.306    8.04    7.51    6.82    5.68    5.25:
              			
HD 250550   15-01-81   44620.627    8.56    7.71    6.83    5.66    5.26
            16-01-81   44621.666    8.64    7.76    6.82    5.63    5.77:
            							     
LkHa 339    01-02-97   50481.442   11.40   10.58    9.70    8.28    --

HD 45677    13-01-81   44618.729    7.65    6.69    4.80    2.16    1.21
            16-01-81   44621.721    7.58    6.74    4.81    2.19    1.28
            							     
LkHa 215    15-01-81   44620.712    8.65    7.97    7.11    5.92    5.47
            16-01-81   44621.783    8.80    8.00    7.33    5.94    5.44:
            							     
HD 259431   15-01-81   44620.718    7.51    6.74    5.76    4.31    3.68
            16-01-81   44621.690    7.54    6.78    5.78    4.35    3.78
            							     
R Mon       15-01-81   44620.726    9.56    8.03    6.12    3.50    2.52
            16-01-81   44621.698    9.54    7.97    6.07    3.41    2.50
            							     
V590 Mon    03-02-97   50483.427   11.77   10.68    9.51    7.74    6.66:

HD 50138    13-01-81   44618.754    5.87    5.14    4.21    2.80    2.18
            16-01-81   44621.730    5.83    5.14    4.20    2.81    2.20
            			 				     
Z CMa       13-01-81   44618.763    5.89    4.88    3.77    1.91    1.08
            14-01-81   44619.774    5.86    4.86    3.76    1.87    1.10
            15-01-81   44620.940    5.81    4.88    3.76    1.90    1.06
            03-02-87   46830        5.83    4.75    3.65    1.64    0.92
            04-02-87   46831        5.83    4.73    3.64    1.64    0.92
            05-02-87   46832        5.82    4.73    3.64    1.64    0.89
            			 				     
HD 53367    13-01-81   44618.775    5.73    5.47    5.22    4.79    4.64
            14-01-81   44619.786    5.70    5.48    5.19    4.74    4.66
            16-01-81   44621.736    5.78    5.48    5.21    4.78    4.59
            			 				     
NX Pup      24-02-86   46486        8.32    7.01    5.82    4.17    3.59
            25-02-86   46487        8.34    7.03    5.83    4.17    3.86
            26-02-86   46488        8.43    7.06    5.84    4.19    3.62
	    
HD 58647    28-01-97   50477.550    8.45    8.47    8.54    9.21:   --      

Hen 3-40    03-02-97   50483.473   12.35   12.04:  12.02    --      -- 

ESO 313-10  01-02-97   50481.547   11.61    9.91    8.60    7.04    5.96:

HD 76534    12-03-84   45772        7.65    7.57    7.49    7.08    --
            13-03-84   45773        7.62    7.54    7.48    7.24    --
	    			  
Herbst 28   01-02-97   50481.580    9.55    9.23    9.08    8.86:   --

HD 87643    13-01-81   44618.819    6.11    5.01    3.77    2.08    1.38
            14-01-81   44619.822    6.05    5.02    3.75    2.03    1.38
            15-01-81   44620.911    6.17    5.01    3.76    2.07    1.37

HD 89249    01-02-97   50481.601    6.70    5.99    5.68    5.37    4.89:   

HR Car      13-01-81   44618.838    6.06    5.65    5.29    4.75    4.38
            14-01-81   44619.832    6.15    5.68    5.30    4.75    4.57
            15-01-81   44621.029    6.19    5.73    5.29    4.74    4.54
            			 				     
Hen 3-480   28-02-91   48315.680   11.20   11.53:  11.69:   --      --
				 
GG Car      13-01-81   44618.860    6.90    6.10    5.09    3.66    3.03
            14-01-81   44619.863    6.85    5.95    4.96    3.63    3.05
            15-01-81   44620.950    6.87    6.04    5.02    3.59    2.97
            			 				     
AG Car      13-01-81   44618.868    4.89    4.53    4.20    3.71    3.56
            14-01-81   44619.871    4.77    4.44    4.07    3.61    3.46
            15-01-81   44620.956    4.89    4.51    4.16    3.70    3.54
            			 				     
CD-76 486   13-01-81   44618.896    7.82    7.05    6.38    5.40    5.28
            14-01-81   44619.798    7.96    --      6.34    5.37    5.33
            15-01-81   44621.010    7.89    7.05    6.33    5.39    5.09
            			 				     
HD 97048    16-02-79   43921        7.15    6.45    5.73    4.66    4.49
            17-02-79   43922        6.88    6.40    5.68    4.69    4.39
            19-02-79   43924        7.08    6.43    5.68    4.61    4.35
            20-02-79   43925        7.16    6.51    5.75    4.68    4.39
            14-01-81   44619.813    6.84    6.85    6.15    4.87    4.62
            15-01-81   44621.016    7.32    6.85    6.13    4.84    4.56
            22-06-83   45508        7.45    6.91    6.23    4.79    4.67:
	    
HD 97300    22-06-83   45508        7.75    7.42    7.19    6.95:   5.6:

HD 100546   04-02-88   47196        6.34    5.75    5.09    4.02    3.75 

HD 101412   10-04-90   47992        8.70    8.24    7.25    5.81    5.08:
	    							    
V644 Cen    29-01-97   50478.619    9.74    9.34    9.04    8.48:   --    

HD 104237   04-02-88   47196        5.76    5.14    4.42    3.05    2.58

DK Cha      28-02-91   48315.866    9.92:   7.49    5.59    3.17    2.51
				 
CD-48 7859  10-04-90   47992       10.04    9.19    7.58    4.62    3.65
            11-04-90   47992.809   10.01    9.17    7.61    4.64    3.66
	    			 
LSS 3027B   25-02-86   46487       12.40:  11.08:  10.55:   --      --
            20-07-86   46632       11.27    9.68    8.35    6.57    5.93:
            20-07-86   46632       11.33    9.68    8.33    6.61    5.77:

HD 130437   03-02-97   50483.622    8.05    7.65    7.28    6.79    7.16::

SS73 44     03-02-97   50483.613   10.18    8.96    7.93    6.55    5.67:

HR 5999     18-05-78   43647        5.78    5.17    4.55    3.13    2.71
            18-05-78   43647        5.76    5.17    4.49    3.12    2.69
            18-05-78   43647        5.70    5.14    4.48    3.18    2.70
            19-05-78   43648        5.77    4.94    4.34    3.02    2.52
            20-05-78   43649        5.63    5.14    4.49    3.02    2.56
            21-05-78   43650        5.57    4.99    4.40    3.03    2.49
            23-05-78   43652        5.64    5.06    4.39    2.83    2.49
            18-02-79   43923        5.79    5.17    4.44    3.28    2.75
            19-02-79   43924        5.66    5.14    4.43    3.38    2.88
            20-02-79   43925        5.66    5.09    4.36    3.27    2.73
            22-06-83   45508        5.89    5.15    4.28    2.88    2.61:
            17-03-84   45777        --      --      4.48    2.97    2.48
            20-07-86   46632        6.08    5.30    4.50    3.13    2.66
	    		      	 				   
HR 6000     18-05-78   43647        6.61    6.76    6.87    --      --
            18-05-78   43647        6.66    6.79    6.85    --      --
            19-05-78   43648        6.93    6.59    6.83    --      --
            20-05-78   43649        6.66    6.74    6.91    --      --
            21-05-78   43650        6.46    6.56    6.80    --      --
            21-05-78   43650        6.52    6.56    6.80    --      --
            23-05-78   43652        6.53    6.67    6.73    --      --
            18-02-79   43923        6.61    6.62    6.62    6.66    6.51:
            20-02-79   43925        6.60    6.54    6.61    6.62    6.14:
	    			 
HD 147196   28-02-91   48315        6.62    6.65    6.50    6.37    6.36
            29-02-91   48316        6.58    6.53    6.47    6.27    --

HD 150193   31-07-85   46278.494    7.02    6.37    5.59    4.30    3.91
	    
AK Sco      20-07-86   46632        7.86    7.25    6.60    5.38    4.98
            21-07-86   46633        7.78    7.20    6.58    5.41    5.07:

CD-42 11721 23-06-83   45509        7.12    5.79    4.37    2.37    1.77
            20-07-86   46632        7.08    5.76    4.39    2.36    1.70
            21-07-86   46633        7.06    5.75    4.37    2.35    1.75
            11-04-90   47993        7.18    5.90    4.45    --      1.64
	    			 
KK Oph      20-06-83   45506        9.25    7.49    5.96    4.02    3.43
	    			 				   
Wra 15-1651 28-02-91   48316.013    --      --      8.36:   6.16    5.41
				 
Wra 15-1678 28-02-91   48315.998    --      --      7.02:   6.26    6.72:
				 
XX Oph      12-05-78   43641        4.15    3.32    3.02    2.77    2.68

HD 163296   09-03-80   44308.222    6.34    5.54    4.72    3.55    3.14
            20-06-83   45506        6.24    5.52    4.70    3.52    3.14
            21-07-86   46633        6.23    5.45    4.54    3.25    2.90
            22-07-86   46634        6.24    5.45    4.56    3.27    2.90

LkHa 112    22-06-83   45508        9.17    8.91    8.55    7.7:    --

LkHa 115    22-06-83   45508       10.6:    9.7:    9.0:    --      --

LkHa 118    22-06-83   45508        8.71    8.26    7.86    7.33:   --  
            12-03-84   45772        8.76    8.26    7.87    7.35    6.69::
            31-07-85   46277.670    8.76    8.31    7.90    7.32    --

LkHa 119    22-06-83   45508        8.32    7.95    7.43    6.50:   6.50::
            13-03-84   45773        9.83    9.44    9.08    --      --
            31-07-85   46277.678    9.91    9.57    9.26    --      --

AS 310      18-05-78   43647        8.12    7.24    6.41    5.22    4.23
            19-05-78   43648        7.89    7.70    6.67    4.63    4.20 
            21-05-78   43650        7.90    6.95    6.26    4.90    4.33
            15-08-79   44101        8.25    7.17    6.22    5.04    4.50

HD 176386   15-08-79   44101        6.93    6.77    6.73    6.58    6.58   
            22-06-83   45508        6.87    6.76    6.68    6.56    6.13::
            31-07-85   46277.759    6.85    6.74    6.66    6.49    6.62   

TY CrA      19-05-78   43648        7.73    7.12    6.91    --      --
            21-05-78   43650        7.57    6.99    6.86    --      --
            15-08-79   44101        7.57    7.05    6.72    6.25    6.27
            22-06-83   45508        7.51    6.97    6.66    6.01    5.85:
            31-07-85   46277.733    7.48    6.97    6.68    6.11    6.16
	    							   
R CrA       18-05-78   43647        7.47    5.70    4.02    1.92    1.37
            19-05-78   43648        7.79    5.84    3.90    2.00    1.28
            21-05-78   43650        7.31    5.61    3.91    1.77    1.07
            15-08-79   44101        7.28    5.27    3.47    1.56    0.93
            22-06-83   45508        7.22    5.09    3.35    1.34    0.84
            31-07-85   46277.792    7.56    5.37    3.64    1.61    1.09

T CrA       18-05-78   43647        --      8.31    7.60    5.8:    5.2:  
            19-05-78   43648        9.40    8.62    7.28    5.29    4.67 
            21-05-78   43650        9.00    8.59    7.15    5.13    4.4:  
            15-08-79   44101       10.42    9.26    --      5.89    4.99
----------------------------------------------------------------------------------
\end{verbatim}

\end{document}